\documentclass[sigconf]{acmart}
\usepackage{multirow}

\AtBeginDocument{%
  \providecommand\BibTeX{{%
    \normalfont B\kern-0.5em{\scshape i\kern-0.25em b}\kern-0.8em\TeX}}}

\setcopyright{acmcopyright}
\copyrightyear{2023}
\acmYear{2023}
\acmDOI{XXXXXXX.XXXXXXX}

\acmConference[Conference acronym 'XX]{Make sure to enter the correct
  conference title from your rights confirmation emai}{September xx-xx,
  2018}{XXXX, NY}
\begin{document}

\title[Exploring the Design and User Experience of In-Vehicle Multi-modal Intuitive Interface]{“Tell me about that church”: Exploring the Design and User Experience of In-Vehicle Multi-modal Intuitive Interface in the Context of Driving Scenario}

\author{Yueteng Yu}
\email{laurence.yu@foxmail.com}
\affiliation{%
  \institution{Tsinghua University}
  \state{Beijing}
  \country{China}
}

\author{Yan Zhang}
\email{yanzhang.hci2022@gmail.com}
\affiliation{%
  \institution{Tsinghua University}
  \state{Beijing}
  \country{China}
}

\author{Gary Burnett}
\email{g.e.burnett@lboro.ac.uk}
\affiliation{%
  \institution{Loughborough University}
  \city{Loughborough}
  \state{Leicestershire}
  \country{United Kingdom}
}

\renewcommand{\shortauthors}{Yu, et al.}

\begin{abstract}
Intuitive interaction has long been seen as a highly user-friendly method. There are attempts to implement intuitive interfaces in vehicles in both research and industrial, such as voice commands. However, there is a lack of exploration in the in-vehicle multi-modal intuitive interaction, especially under a dynamic driving scenario. In this research, we conducted a design workshop (N=6) to understand user's needs and designers' considerations on the in-vehicle multi-modal intuitive interface, based on which we implemented our design on both a simulator and a real autonomous vehicle using Wizard-of-Oz. We conducted a user experiment (N=12) on the simulator to explore determinants of users' acceptance, experience, and behavior. We figured that acceptance was significantly influenced by six determinants. Drivers' behavior has an obvious pattern of change. Drivers have been proven to have less workload but distractions were also reported. Our findings offered empirical evidence which could give insights into future vehicle design.
\end{abstract}




\begin{teaserfigure}
  \centering{\includegraphics[width=1\textwidth]{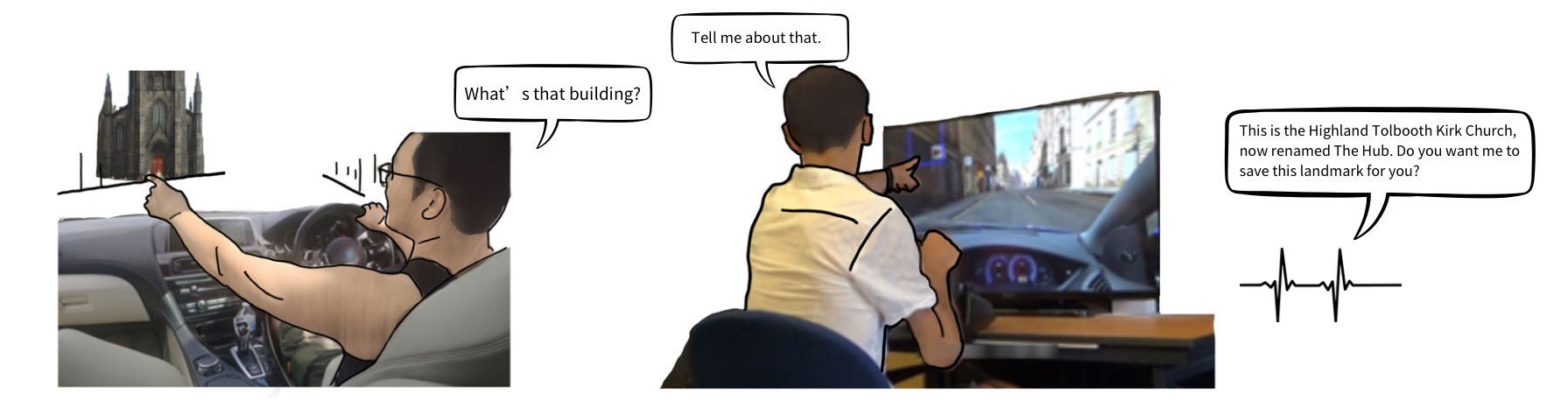}}
  \caption{An overview of the designed in-vehicle multi-modal intuitive interface. The scene reproduction of motivation (left) and the design interface solution (right).}
  \label{fig:teaser}
\end{teaserfigure}

\received{20 February 2007}
\received[revised]{12 March 2009}
\received[accepted]{5 June 2009}

\maketitle

\section{Introduction}
The interactive technique between the driver and the in-vehicle intelligent system is becoming an increasingly popular topic in the automotive area \cite{invehicleTechniReview}. 
The usage of intuitive interaction has been explored considering its simplicity and low workload requirement for drivers \cite{riener2011natural, riener2011natural2, gui2022going}. 
Various intuitive modalities have proven potential, such as speech, visual display, eye gaze, gesture, etc. 
A multi-modal system that processes two or more combined modalities enables an efficient, flexible, and expressive user experience \cite{oviatt2007multimodal}. It is consistent with and can be enhanced by intuitive interaction, which has advantages in providing non-consciously usage with users' prior knowledge without the effort of learning with high user acceptance \cite{blackler2006towards, blackler2005intuitive, mohs2006iuui}. 

Voice-user interface (VUI) with intelligent dialogue, has been widely used for in-vehicle commands, such as navigation, without glancing or touching the physical screen \cite{zheng2017navigationorientatedvoiceassistant}. A comprehensive review showed the speech assistant could reduce distractions by keeping drivers' eyes on the road and hands on the steering wheel \cite{invehicleTechniReview}. Augmented Reality (AR) is a technology that adds computer-generated information to the view of the physical world through an augmented environment \cite{carmigniani2011augmentedARreview}.
AR head-up displays (HUDs) have been designed to fuse virtual elements with the real world to help drivers maneuver and demonstrate additional information \cite{invehicleTechniReview, gabbard2014behindtheglass}. Full windshield HUD (FW-HUD) was proposed and evaluated to provide drivers with a more immersive experience \cite{hanel2016towardswindshielddepthcalculation, park14projectorbasedwindsheild, topliss19evaluatinghud, marevska22production}. Another novel in-vehicle input modality, deictic gestures, has been explored recently by major automotive original equipment manufacturers (OEMs). Human's deictic gesture, which occurs before the infant’s first birthday \cite{capone2004gesture}, is a natural intuitive communication to indicate objects referencing the current environment. Therefore, it is a considerable intuitive input method.

However, there is a lack of in-depth discussion on the design of in-vehicle multi-modal intuitive interaction and its user experience, especially in the context of dynamic driving scenarios. 
Moreover, the above-mentioned literature is mostly limited to information and actions inside the cockpit. 
The joint inside-outside vehicle information exchange needs more exploration.

In this research, our contributions to fulfill the gaps can be concluded as i) A design workshop to explore users' needs and potential solutions for an in-vehicle multi-modal intuitive interface.
ii) Implementations of the designed interface on both a driving simulator and a real autonomous vehicle.
iii) A user experiment in the simulator to evaluate users' acceptance, experience, and behavior in a driving scenario.

\section{Related Works}
\subsection{The Development of In-Vehicle Intuitive Interface}
The design of an intuitive interface has been explored by researchers in various human-computer interaction (HCI) areas, such as robotics \cite{villani2018survey}, education \cite{lai2018learning}, and accessibility \cite{choi2008laser}, as it has low learning cost, low workload, and high user experience \cite{blackler2007towards}. For private cars, as machines with a wide consumer audience \cite{handy2002accessibility}, it is essential to equip them with an intuitive interface that can facilitate diverse groups of users. Various types of intuitive interaction have been used in vehicles. 

\subsubsection{Voice-User Interface}
Voice assistants are commonly implemented for OEMs (e.g., Daimler's MBUX, BMW Intelligent Personal Assistant) and also integrate third-party assistants such as Amazon Alexa \cite{invehicleTechniReview}. Navigation is one of the popular features that use VUI to command without glancing at or touching the physical screen, therefore safety. Considerable works have designed and developed intelligent vehicle dialogue. For instance, \cite{zheng2017navigationorientatedvoiceassistant} designed a Deep Neutral Network (DNN) based on Natural Language Processing (NLP) to improve the vehicle's human-machine interface (HMI) performance. Moreover, it was suggested that passive listening might not be as effective as active conversation when driving \cite{invehicleTechniReview}. Studies showed that more natural and intuitive communications could prevent drivers' distraction \cite{large2018drivenVUI1, large2019lessonsVUI2}. 

\subsubsection{Head-UP Displays}
Considerable studies have suggested that AR HUD is a technology that can prevent drivers from taking their eyes off the road \cite{2017HUDLaneChanges, smith2021isolatingHDDHUD}. Furthermore, by combining the location-based services, an interaction that fused the user input and 3D HUD was studied to allow drivers to obtain additional information while minimizing cognitive and visual distraction \cite{DriverQueries3dcursor}.

\subsubsection{Gesture Controls}
Baudel and Beaudouin-Lafon \cite{Baudel1993RemoteFreeHandGestures} emphasized the possibility and potential of the remote free-hand pointing gesture by demonstrating a DataGlove to control the application. It was suggested that using free-hand gestures for interaction has the advantages of being "natural", "terse and powerful" and "direct", which consists of intuitive actions. Likewise, the naturalness of the advantage was also presented by the pioneering research of deictic gesture \cite{putthatthere}.

Nowadays, gesture-based instructions have been developed and implemented in a variety of fields. In the AR scenario, users can use the pointing gesture to achieve basic functionalities of menu control, object referencing and object learning \cite{AR2004pointinggesture}. For robotics, considerable evidence has attempted to track and recognize the pointing gesture in a 3D environment \cite{nickel20043dsdtrackingpointinggesture, droeschel2011learningLRdepthCam1}. Moreover, a majority of gesture tracking studies have investigated how gesture controls can help us complete our daily tasks more efficiently \cite{InAirGestures2014MobileDevices, ultraleaphaptic2018exploring}. For example, VisionWand allowed users to engage with remote places on a huge display by pointing with a smart smartphone \cite{CaoLargeDisplays2004pointingsmartphone}.

As for the cockpit interface, it is shown that in-vehicle gesture control is safer than typical touch and tactile interaction by reducing the demands on driving attention \cite{ba2008you, riener2012gestural}. Moreover, \cite{stopoverthere} suggested the combination of pointing gesture and speech required less cognitive demand than speech and touch in an autonomous driving scenario. However, only a few works \cite{Aftab2022multimodalinsideoutside, gomaa2020studyingOutsidePointMovingCar, youhaveapointthere2020insideobject, DriverQueries3dcursor, sauras2017voge, stopoverthere, freehandpoint} investigate the in-vehicle deictic interface, which is closely related to driving scenarios, sensors, and algorithms in the car.

\begin{table*}[]
\caption{The summary of in-vehicle multi-modal intuitive interfaces with deictic gestures.}
\label{tab:RLtable}
\resizebox{\linewidth}{28mm}{
\begin{tabular}{|c|c|c|c|c|}
\hline
Ref & Modality Fusion & W/WO Driving & Scenario & Tasks \\ \hline
This study & \begin{tabular}[c]{@{}c@{}}Speech (input \& output), \\ Pointing (outside), HUD\end{tabular} & Yes & Travel & Point at interest targets \\ \hline
\cite{stopoverthere} & \begin{tabular}[c]{@{}c@{}}Speech (input), Pointing (outside), \\ Touch (input)\end{tabular} & No & Semi-immersive VR & \begin{tabular}[c]{@{}c@{}}Parking, waving friends, \\ highway exit, lateral swift\end{tabular} \\ \hline
\cite{sauras2017voge} & Speech (input), Pointing (outside) & N/A & Prototype & N/A \\ \hline
\cite{freehandpoint} & \begin{tabular}[c]{@{}c@{}}Pointing (inside \& outside), \\ Eye gaze (input), Head (input)\end{tabular} & No & Sightseeing & Point at interest targets \\ \hline
\cite{DriverQueries3dcursor} & Pointing (outside), HUD (3D) & No & Recorded scenes & Pointing at fixed targets \\ \hline
\cite{gomaa2020studyingOutsidePointMovingCar} & Pointing (outside), Eye gaze (input) & Yes & Pure driving & Pointing at fixed targets \\ \hline
\cite{Aftab2022multimodalinsideoutside} & \begin{tabular}[c]{@{}c@{}}Pointing (inside \& outside), \\ Eye gaze (input), Head (input)\end{tabular} & No & Arounded by buildings & Pointing at fixed targets \\ \hline
\cite{youhaveapointthere2020insideobject} &\begin{tabular}[c]{@{}c@{}}Pointing (inside), \\ Eye gaze (input), Head (input)\end{tabular} & No & In cockpit & Pointing objects in cockpit \\ \hline
\end{tabular}
}
\end{table*}

\subsubsection{Other Modalities}
Other intuitive interaction control within vehicles includes the brain-computer interface (BCI), which is still in the relatively early stages of research. For instance, there have been studies attempting to implement BCI-based control of car functions in driving simulators \cite{Hood2012} and real smart vehicles \cite{Nianming2023}. \cite{Daniel2013} introduced a semi-autonomous car control method based on a BCI, exploring the use of four different brain patterns to control the steering wheel and throttle/brake. However, vehicle control through BCIs continues to face various challenges, including safety and ethical considerations.

Research into driver safety monitoring has also extensively explored gaze-based detection \cite{rozanowski2012infrared, barbuceanu2009eye}. For example, Han introduced a novel approach that incorporates a driver's dynamic gaze attention into the interaction within the human-vehicle system \cite{Han2022}. This study aimed to dynamically adjust the angle of the rear-view mirror based on the driver's area of interest (AOI). Gomaa et al \cite{gomaa2020studyingOutsidePointMovingCar} designed a system that used eye gaze to select outside-the-vehicle objects. The average total gaze duration was 5.04 seconds.
Additionally, touchless Biometric Systems have been under investigation in the context of vehicle interaction \cite{Banerjee2020}. 

However, due to the technology limitations and driving safety, such as eyes off the road, the BCI, eye-tracking, and other biometric detections were not considered in this research.

To conclude, the goal of in-vehicle intuitive interaction is to increase safety by reducing the cognitive workload, and the demand for visual attention \cite{invehicleTechniReview}. In this study, we further discussed the possible design solutions for in-vehicle intuitive interfaces with different modalities and application scenarios.

\subsection{Designs for In-Vehicle Multi-modal Interface}
The multi-modal interface has been recognized for being intuitive, robust, and preferred \cite{naumann2009multimodal, oviatt1999ten} as it is comparable to natural human-human communications \cite{chen2006designing}.

Researchers are exploring the advantages of fusing modalities for in-vehicle interfaces. A comprehensive summary has been made (see table \ref{tab:RLtable} to conclude the literature studying in-vehicle multi-modal interactions. As deictic gesture interaction is an effective intuitive behavior yet lacks discussion, only works with modalities including pointing gestures have been chosen. It is noticeable that most designs were implemented and evaluated under a static scenario without driving, which can cause inaccuracies as most usage scenarios are dynamic.

In this study, we combine three intuitive modalities under a dynamic traveling scenario. A design workshop has been organized to explore users' needs and the interface design solutions. The system has been implemented on both a driving simulator and a real autonomous vehicle. Critical findings have been extracted from the user experiment.

\section{Study One: Design Workshop}
To understand users' needs and explore potential solutions for the in-vehicle multi-modal intuitive interface, a design workshop has been held.

\subsection{Methods}
\subsubsection{Participants}
Table \ref{table:wsparticipant} shows participants' demographic information ($M_{age} = 23.00$, $SD_{age} = 1.73$; $Male:Female = 1:1$). 50\% of them have an interaction design background and they have self-rated their degree of expertise ($M = 3.50$, $SD = 1.26$). They are all licensed drivers ($M = 2.50$, $SD = 0.50$). They were grouped in two to cooperate and discuss in the workshop.

\begin{table}[]
\caption{The participants' demographic of the design workshop.}
\label{table:wsparticipant}
\begin{tabular}{ccccc}
\hline
PID & Age & Gender & \begin{tabular}[c]{@{}c@{}}Driving\\ Experience (yr)\end{tabular} & \begin{tabular}[c]{@{}c@{}}Degree of expertise \\ in interaction design (1-5)\end{tabular} \\ \hline
1 & 21 & Female & 2 & 5 \\
2 & 25 & Female & 3 & 4 \\
3 & 24 & Male & 3 & 2 \\
4 & 22 & Male & 2 & 3 \\
5 & 21 & Female & 2 & 2 \\
6 & 25 & Male & 3 & 5 \\ \hline
\end{tabular}
\end{table}

\subsubsection{Procedures}
The workshop had four main sections with a duration of 1.5 hours. At the beginning, an experimenter provided background information to participants, which included the definition of intuitive interaction and examples of the current state-of-the-art in-vehicle interface.

In the second section, participants were asked to imagine they were User Experience designers of automobile manufacturers. They were organized to brainstorm the possible modalities for future in-vehicle intuitive interaction, the functions, usage scenarios, and combinations of different modalities. Their ideas were written on stickers simultaneously.

In the third section, a persona and a scenario have been made based on the result of the previous brainstorm. 
\begin{quote}
    Allen is a young man who loves road trips. One day, he was driving in a city for the first time. While driving around, he saw a medieval-style building and was curious about its name and history. However, he didn't want to pull over so he couldn't search for its information in time. He had to pass with regret.
    Are there any in-vehicle interface design that could help Allen so he could learn about interesting building next time?
\end{quote}
Participants discussed the design solution and visualized their idea on a paper. Their solutions were presented and discussed together.

The last section was a semi-structured interview. Some questions were pre-designed and some were generated based on the results of previous discussions. The materials can be found in the Appendix \ref{ap:workshop}.

\subsubsection{Data Analysis}
The brainstorming, design presentation, and semi-structured interview transcripts were coded and analyzed by Grounded Theory \cite{white2022grounded}. The stickers collected from the brainstorm generated an Affinity Diagram \cite{lucero2015using} by two researchers to identify users' needs and designers' ideas for future interfaces. 

\subsection{Results}
\subsubsection{Design Considerations}
\begin{figure*}
    \centering
    \includegraphics[width=1\textwidth]{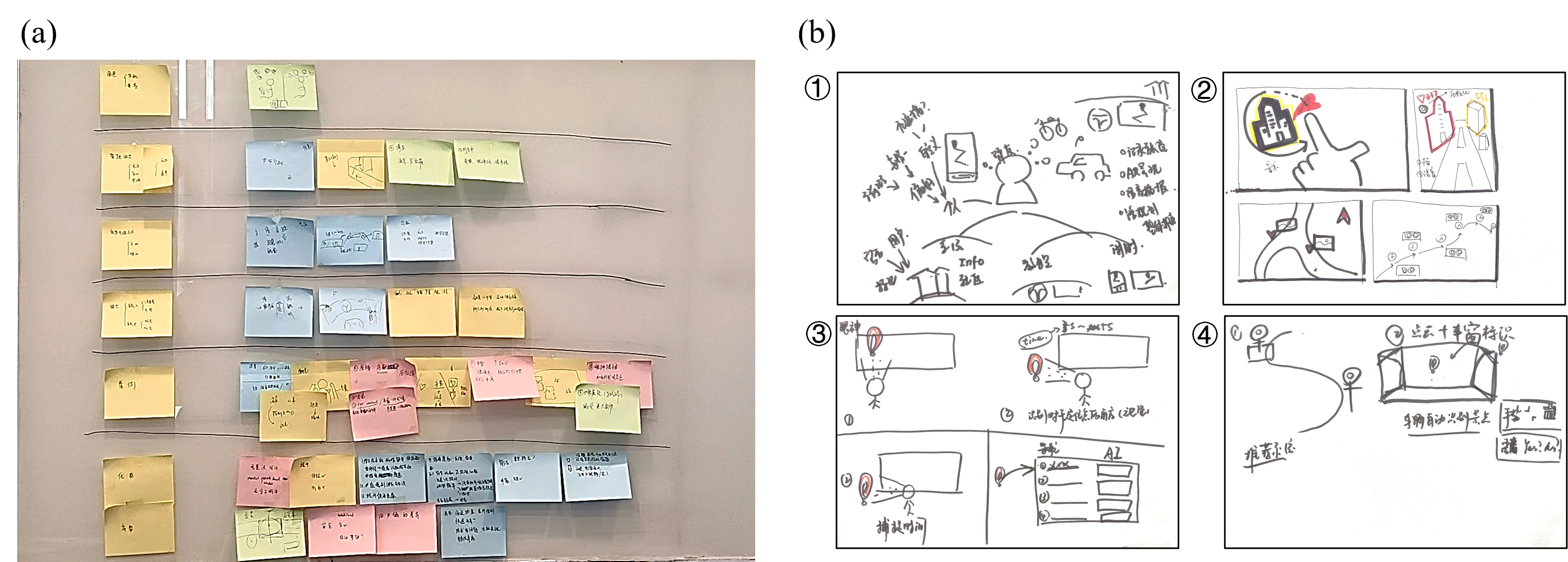}
    \caption {The result of the design workshop. (a) The affinity diagram. (b) Participants design solutions.}
    \label{fig:workshop}
\end{figure*}

The transcripts of the workshop have been conceptualized into five parts, actors, modalities, way of information exchange, driving status, and pros and cons (figure \ref{fig:workshop}a).

Actors as been categorized into drivers and passengers. In this discussion, participants tacitly mainly discussed drivers.
Modalities included audio, visual, gesture, and tactile, which could be combined in different ways. 
The way of information exchange consisted of active and passive ways. 
The driving status has been categorized into dynamic and static states. The dynamic state included "focused on driving" and "distracted by secondary tasks". The static states stood for drivers' idle time without driving. 

During the dynamic states, drivers may use less active ways to exchange information but more passive ways, such as HUD for visual information. Speech input, as an active way, however, requires less mental workload, which can be the input modality for dynamic states. When the traffic condition is simple and safe, drivers may carry out secondary tasks. Therefore, more modalities can be used as active inputs, such as gestures.
The following part shows examples of designed functions under different scenarios, which will be discussed according to their categories.

\begin{itemize}
    \item P6 described a scenario when the driver is driving on the road (\textit{dynamic states}). The driver sees an interesting building and gives a finger-heart gesture (\textit{gesture modality, active way}) pointing at the building. Next time, when the driver passes by, the heart icon will show on the windshield in AR-style (\textit{visual modality, passive way}).
    \item P6 provided another scenario when drivers are experiencing an extreme traffic jam. Drivers are waiting in queues (\textit{static state}) and feel irritable. There can be an AR chat room showing on the windshield (\textit{visual modality, passive way}) where nearby drivers can chat (\textit{audio modality, active way}) to pass the time. 
    \item P3 demonstrated a scenario when a driver is traveling on the road at a low speed (\textit{dynamic state}). The driver may miss some places of interest because of the unfamiliarity. It can be a convenient way to demonstrate photos and information of surrounding famous places on AR HUD (\textit{visual modality, passive way}), by localizing the car using GPS.
    \item P2 gave some examples of optimizing existing functions. Drivers can use pointing gestures and knob gestures (\textit{gesture modality, active way}) to control sound and air conditioning. Seats can be adjusted to the most comfortable way for different people using pressure sensing and tactile feedback (\textit{tactile modality, passive way}).
\end{itemize}

The pros and cons of the multi-modal intuitive interface have been discussed by participants. The most mentioned advantage was the user's acceptance level. According to P6 and P1, as the intuitive interaction follows users' mental modal, the learning cost is low enough for users to accept a new interface very soon. Meanwhile, it is worth noticing that intuitive interaction can improve the experience but not change driving behavior. The high degree of usability and entertainment were the other two mentioned advantages. Moreover, multi-modalities interaction was a popular solution in this workshop. P3 and P6 agreed that different modalities can be complementary.

However, there were several possible disadvantages and limitations of the in-vehicle intuitive interface. The most concerned one was safety, as interaction with the interface can be considered as a non-driving related task that causes dangerous distractions. The increment in workload was also a risk to safety. Although it can be lower compared with other types of interface, it still cannot be ignored. Participants mentioned that autonomous vehicles would be a suitable solution. Moreover, the development of technology is another important limitation of the interface. P4 argued the accuracy of intuitive input detection can strongly affect user experience.

\subsubsection{Designed Interface}
Based on the discussion in brainstorming, researchers created a persona "Allen" and a scenario "travel" for participants to design a future interface. Figure \ref{fig:workshop}b shows the visualized results of different groups (3 and 4 were from the same group).

The first group used the idea of crowdsourcing to collect drivers' preferences on interesting buildings. During Allen's driving, the system would highlight the recommended building on the AR windshield and provide audio information. If Allen accepts the recommendation, the system can provide route navigation to the building. The driver can also save this location for a later visit.

The second group provided an idea using pointing gestures. If the driver points at a building along the road, the HUD will highlight the outlines of the target and show relevant information. The driver can also give speech commands as an aid. After marking the building, the car will take a photo of the view and save it into maps. Allen can review the buildings he has passed through after the trip.

The third group used eye-tracking and head-tracking rather than pointing gestures to indicate interested targets. After capturing the building, the system will recognize and add this building to the user's navigation list and recommend similar buildings later on. Moreover, the recommended building will be marked on the windshield. If it does not match the user's preference, a gesture of zoom out and a speech command of negative feedback can help to improve the system.

The configurations of visual feedback that highlights the building and audio feedback to provide relevant information have been discussed further. P6 believed a square pointer could be a clear indication of the target building. However, P1 and P2 argued the arrow pointer could address the situation that is unable to detect a specific target. Also, they believe there can be more possibilities to implement special functions on the arrow pointer, such as indicating a target that is not in the field of view. P3 thought there was no difference in different pointers. For the audio feedback, all participants agreed whether the speech information should be rich or simple depends on different situations. There was no monopoly choice.

\subsubsection{Summary}
In this workshop, we discovered users' needs and explored potential design solutions for the intuitive interface, especially under a road trip scenario. Pointing gestures and speech commands were important intuitive inputs. Visual feedback and audio information were preferred outputs to provide information. Other functions, such as saving the target location, have been considered necessary in the system. Overall, the in-vehicle multi-modal intuitive interface can be concluded as a user-friendly, high-acceptance, and low-learning cost method that can benefit users in the future. There are also limitations, such as safety and technical realization, which may be tackled by the rapid development of autonomous driving.

\section{System Design}
\begin{figure*}[h]
    \centering
    \includegraphics[width=1\textwidth]{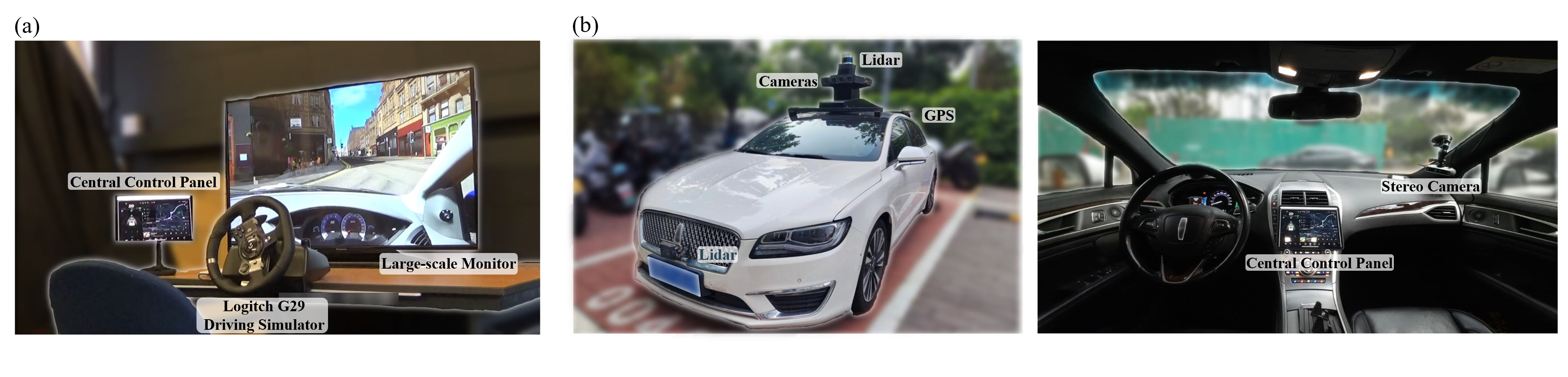}
    \caption{Interface implementation. (a) The system is set up in the driving simulator. (b) The system is set up in a real autonomous vehicle.}
    \label{fig:setup}
\end{figure*}

We designed a multi-modal intuitive interaction system that complies with the findings from the design workshop, in both a driving simulator and on a real autonomous vehicle (see figure \ref{fig:setup}). 
Referring to the system design from the multi-modality automotive user interfaces \cite{freehandpoint, stopoverthere, Aftab2022multimodalinsideoutside} and from the previous workshop, speech control, deictic gesture recognition, and windshield visual support were added. Hence, the intuitive interface in this study included three components: pointing gesture input, voice input and response, and pointer display.

\subsection{In Simulator}
Following the system setup design by \cite{freehandpoint}, our system imitated an actual car cockpit that has multiple interactive sensors and devices to create a similar experience. As shown in figure \ref{fig:setup}a, the Logitch G29 driving wheel and pedals were used as the hardware of a driving simulator. The controller setting was modified to the simulation parameters to create a real car driving experience rather than a racing game. A large-scale monitor simulated the windshield displayed the driving game \textit{Forza Horizon 4} with high-fidelity graphics. The right-hand drive Ford Focus RS was used to simulate driving conditions in the UK. A tablet was mounted on the left-hand side to simulate the central control panel, which provided HD maps and speech assistance.

The intuitive interface would recognize users' pointing gestures to the buildings shown on the monitor and provide feedback by displaying 2D pointers on the screen to highlight the target building. After users gave deictic voice commands, such as ”Tell me about \textit{that} building.", the speech assistant would react with appropriate voice responses to provide information. Navigation information and an animation of voice are shown on the central control panel. To support the following user experiment, the system was realized the by Wizard-of-Oz method \cite{DAHLBACK1993wizardOfOz}, which will be described in detail in the User Experiment section.

\subsection{On Real Autonomous Vehicle}
The same design has been implemented on a real autonomous vehicle (AV), which was a non-commercial customized vehicle (figure \ref{fig:setup}b left). By taking advantage of the existing sensor hardware of AVs, implementing an inside-outside multi-modal interface would be more convenient \cite{stopoverthere}. Moreover, as P4 said in the design workshop, autonomous vehicles could be a safer solution for this interface design.

For the multi-modal intuitive interface (figure \ref{fig:setup}b right), a real-sense D455 camera was mounted on the side to support gesture recognition, which could be united with RGB cameras outside the vehicle to realize target object detection. The central control panel with voice assistance can provide command recognition and target information with integrated Google Maps. Due to the limitations of current techniques, only a few solutions for visual displays on the windshield could be found. Thus, the pointer display modality remained for future development.

\section{Study Two: User Experiment}
To evaluate users' acceptance, experience, and behavior of the designed multi-modal intuitive interaction system, we conducted a user study on the simulator. The system on the real autonomous vehicle was not examined because of ethics and safety considerations. The ethics regulation was approved by the university and consent was received from all participants.

\subsection{Method}
\subsubsection{Approach}
\begin{figure*}
    \centering
    \includegraphics[width=0.6\textwidth]{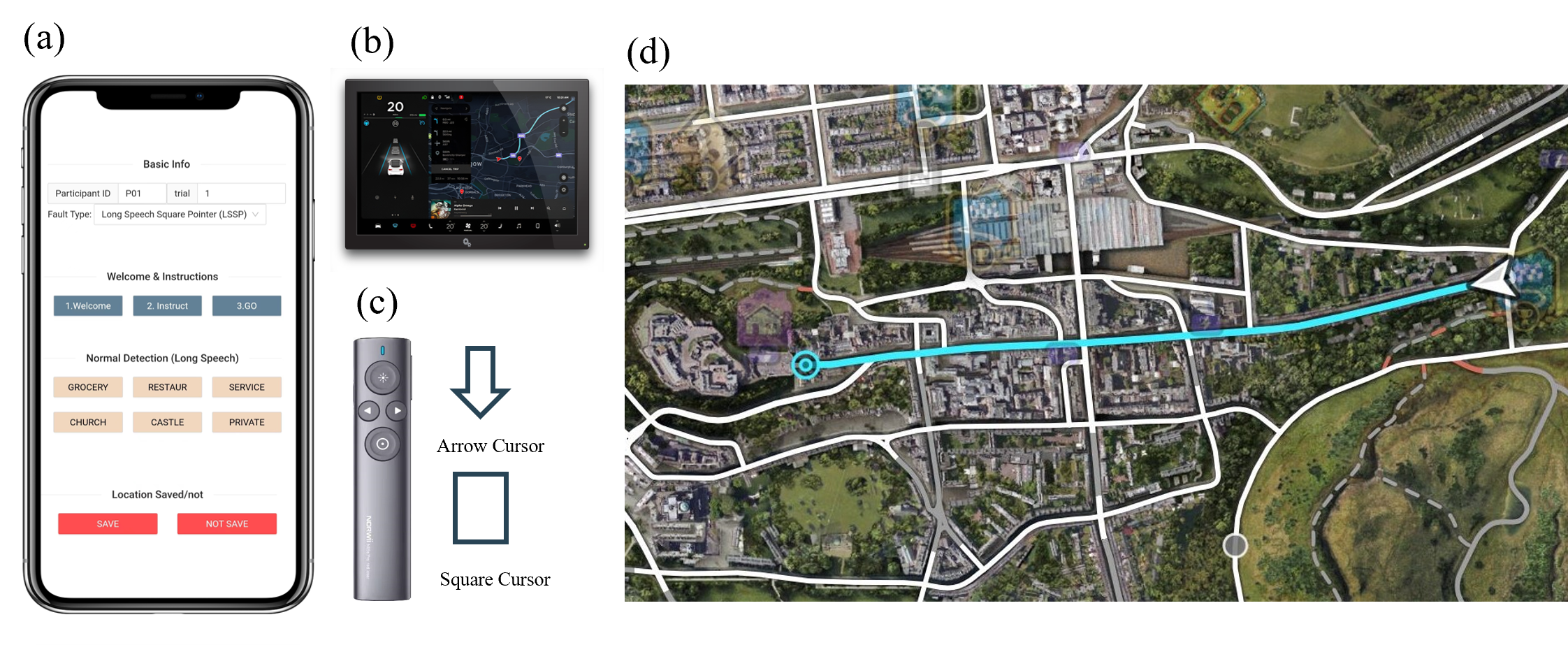}
    \caption{The experiment materials. (a) The experimenter speech control tool kit. (b) The central control panel of the simulator. (c) The Norwii presenter and customized pointer. (d) The route that was chosen for this study.}
    \label{fig:materials}
\end{figure*}

As mentioned in the System Design section, the Wizard-of-Oz method was used discreetly. An experimenter acted as an intelligent system to precept and process users' input and give appropriate feedback. As Mayer \cite{ImprovingHumansInterpretDeicticMayer20} suggested the rotation and distance between the experimenter and the pointer (driver) affect the accuracy of interpreting the deictic gestures, the experimenter stood behind (180°) the participant as close as possible without their knowledge, by pretending to take notes for their driving behavior observation. The experimenter judged which object the subject was pointing at and projected the customized 2D pointer onto the screen with a Norwii presenter to provide visual feedback. The audio interface implemented on the central control panel was triggered by the experimenter using a self-developed control tool kit on a smartphone (shown in figure \ref{fig:materials}). The tool kit was developed by React.js, which included a Basic Information recording section, a Welcome \& Instructions section, buttons of all the Landmarks that participants might point at during the experiment to trigger speech response, and a landmark Saving section.

A travel scenario in Edinburgh was chosen to motivate participants' deictic actions. Participants were asked to act as self-driving tourists who were trying to find some interesting places along the road. The selected route for this experiment exists in the real world. Landmarks that drivers could see were castles, churches, a variety of gift shops, and restaurants. A straight road from a roundabout to Edinburgh Castle was chosen (see figure \ref{fig:materials}d), on which there were 201 valued landmarks that might attract drivers' attention \cite{burnett2001supportinggoodlandmarks}. The summary and descriptions of the landmarks are shown in table \ref{table:landmarks}. To enable the audio modality of the interface, 58 audios of landmark information, 12 instruction, and general response audios were prepared by an AI voice generation platform, which were triggered by experimenters. A sample of speech content can be found in table \ref{table:landmarks}.

\begin{table*}[]
\caption{Summary and descriptions of landmarks}
\label{table:landmarks}
\begin{tabular}{cccl}
\hline
Landmark & Count & \begin{tabular}[c]{@{}c@{}}Number of\\ Audios\end{tabular} & \multicolumn{1}{c}{Samples of Speech Contents} \\ \hline
\multirow{2}{*}{Castle} & \multirow{2}{*}{1} & \multirow{2}{*}{2} & \begin{tabular}[c]{@{}l@{}}Long: Oh, that’s Edinburgh Castle... One of the oldest fortified places\\ in Europe. 5 stars rated in this city. Do you want me to save the location\\ for you?\end{tabular} \\
 &  &  & Short: Oh, that’s Edinburgh Castle... Do you want me to save this landmark? \\ \hline
\multirow{2}{*}{Church} & \multirow{2}{*}{4} & \multirow{2}{*}{6} & \begin{tabular}[c]{@{}l@{}}Long: This is the Highland Tolbooth Kirk Church, now renamed The Hub.\\ Do you want me to save this landmark for you? \end{tabular} \\
 &  &  & Short: This is the Highland Tolbooth Kirk Church. Want to save it? \\ \hline
\multirow{2}{*}{Shop} & \multirow{2}{*}{67} & \multirow{2}{*}{20} & \begin{tabular}[c]{@{}l@{}}Long: This is DJ Shop, 4.3 stars rated on Google map. Do you want me to\\ save the location for you?\end{tabular} \\
 &  &  & Short: This is DJ Shop. Do you want to save it? \\ \hline
\multirow{2}{*}{Restaurant} & \multirow{2}{*}{86} & \multirow{2}{*}{20} & \begin{tabular}[c]{@{}l@{}}Long: The place you pointed at is Treacle. A 4.1 stars Indian restaurant\\ rated on Google map. It is open until 1 AM. Do you want me to save \\ the location?\end{tabular} \\
 &  &  & Short: This is Treacle Indian restaurant. Do you want to save it? \\ \hline
\multirow{2}{*}{\begin{tabular}[c]{@{}c@{}}Private\\ House\end{tabular}} & \multirow{2}{*}{9} & \multirow{2}{*}{6} & \textit{Long}: Sorry, no information is found based on the place you are pointing at. \\
 &  &  & \textit{Short}: Sorry, I can’t find any information. \\ \hline
\multirow{2}{*}{Service} & \multirow{2}{*}{34} & \multirow{2}{*}{4} & \begin{tabular}[c]{@{}l@{}}Long: This place is available to rent, didn’t found any information\\ on the Google map.\end{tabular} \\
 &  &  & Short: This place is closed now. \\ \hline
Total & 201 & 58 &  \\ \hline
\end{tabular}
\end{table*}

\subsubsection{Participants}
Twelve participants ($M_{age} = 27.42$, $SD_{age} = 5.69$, $Male:Female = 1:1$) with driving licenses or driving game experience ($M_{experience}=3.42$, $SD_{experience} = 5.01$) were recruited from the university (see table \ref{table:participants}). Four out of twelve participants had experience with gesture control, such as VR games, switch joysticks, and interaction devices in galleries. Nine out of twelve participants had experience with voice control, most of which were personal voice assistants, such as Siri and Alexa. Two participants used in-vehicle voice control before, one for windows and sunroof control, the other for navigation commands.

\begin{table*}[]
\caption{Experiment participants' demographics. (*: for windows and sunroof control; **: for navigation)}
\label{table:participants}
\begin{tabular}{cccccc}
\hline
PID & Age & Gender & \begin{tabular}[c]{@{}c@{}}Driving\\ Experience (yr)\end{tabular} & \begin{tabular}[c]{@{}c@{}}Voice Control\\ Experience\end{tabular} & \begin{tabular}[c]{@{}c@{}}Gesture Control\\ Experience\end{tabular} \\ \hline
7 & 28 & Male & <0.5 & N/A & N/A \\
8 & 25 & Male & 0.5 & Siri & VR \\
9 & 24 & Female & 1 & Siri & VR \\
10 & 30 & Male & 6 & Amazon Alexa & Switch \\
11 & 28 & Female & 4.5 & In-vehicle* & N/A \\
12 & 25 & Female & 1 & N/A & N/A \\
13 & 28 & Female & 2.5 & N/A & N/A \\
14 & 45 & Male & 19 & Google Assistant and Alexa & N/A \\
15 & 25 & Female & 3 & Google Assistant & N/A \\
16 & 24 & Female & <0.5 & Siri and In-vehicle** & N/A \\
17 & 23 & Male & 2.5 & Siri & N/A \\
18 & 24 & Male & 1 & Siri & Gallery \\ \hline
\end{tabular}
\end{table*}

\subsubsection{Procedures}
After signing the consent form, participants were introduced to the intuitive interface and driving scenario. They were asked to imagine having a road trip to a new city and finding some interesting places along the road. Before the experiment started, there was a training session for participants to get familiar with the simulator for 5 minutes after a briefing of the hardware and software. 

Each participant completed four trials of driving with a counterbalanced order, which were Long Speech and Square Pointer, Long Speech and Arrow Pointer, Short Speech and Square Pointer, Short Speech and Arrow Pointer. According to results from our workshop, users have a similar preference for Long Speech \& Short Speech and Square Pointer \& Arrow Pointer, we provided all the combinations of the speech and pointer configurations for participants to experience but not for a comparison purpose. During each trial, participants need to interact with the in-vehicle interface and save/not save their interested landmarks. Each trial took about 4 minutes with a break time in between to avoid simulation sickness. 

After each trial, the Technology Acceptance Model 3 (TAM3) \cite{venkatesh2008tam3}, Subjective Workload Assessment Technique (SWAT) Index \cite{REID1988swatworkload}, 5-point Likert scales of speech and pointer configurations preference, and an semi-structured interview were taken. All the driving videos and vehicle telemetry data were recorded during the experiment. The details of the interview can be found in Appendix \ref{ap:exp}.

\subsubsection{Evaluations}
\begin{enumerate}
\item Acceptance: To analyze users' adoption and usage of the designed intuitive interface, TAM3 has been used. The correlation analysis of each determinant has been conducted. Relationships of significant Pearson correlation of TAM3 factors have been calculated.

\item Hybrid Analysis for Behaviors: To analyze drivers' driving and interacting behavior, we conducted a hybrid analysis on both their driving videos (for qualitative analysis) and vehicle telemetry data (for quantitative analysis). The driving videos were manually coded using Boris \footnote{https://www.boris.unito.it/}, aiming to identify events. 

\item Workload: To reveal drivers' workload when using the intuitive interface as a secondary task of driving, SWAT has been used and analyzed.

\item Preference: To verify the result from the previous workshop that users have similar preferences for different speech and pointer configurations, the scores of Likert Scales have been counted.

\item Subjective feelings: A semi-structured interview was conducted to explore participants' subjective feelings about our system, such as advantages, disadvantages, limitations, and possible improvements.
\end{enumerate}

\subsection{Results}
A total of 12 participants completed the experiment and 48 trials of valid results were collected. An average of 5.16 pointing behavior for each trial has been recorded.

\subsubsection{User Acceptance}
A descriptive statistics analysis on the scores of TAM3 determinants has been conducted, the result is shown in table \ref{table:tam}. The Perceived Ease of Use obtained the highest score ($M = 4.29$) and Computer Anxiety obtained the lowest score ($M = 2.83$).

\begin{table*}[]
\caption{The descriptive statistics results and correlation analysis of determinants for TAM3. (*: p<0.05, **: p<0.01)}
\label{table:tam}
\begin{tabular}{ccccc}
\hline
\multirow{2}{*}{Determinants} & \multirow{2}{*}{Mean (Std)} & \multicolumn{3}{c}{Pearson Correlation (Sig. (2-tailed))} \\ \cline{3-5} 
 &  & \begin{tabular}[c]{@{}c@{}}Behavioral\\ Intention (BI)\end{tabular} & \begin{tabular}[c]{@{}c@{}}Perceived\\ Usefulness (PU)\end{tabular} & \begin{tabular}[c]{@{}c@{}}Perceived Ease\\ of Use (PEOU)\end{tabular} \\ \hline
Behavioral Intention (BI) & 3.63 (0.65) & - & - & - \\
Perceived Usefulness (PU) & 4.10 (0.69) & 0.638 (0.017*) & - & - \\
Perceived Ease of Use (PEOU) & 4.29 (0.57) & 0.623 (0.015*) & 0.74 (0.003**) & - \\
Subjective Norm (SN) & 3.40 (0.67) & 0.435 (0.079) & 0.026 (0.468) & - \\
Image (I) & 3.04 (0.72) & -0.277 (0.191) & -0.285 (0.185) & - \\
Job Relevance (JR) & 4.17 (0.67) & 0.586 (0.023*) & 0.702 (0.005**) & - \\
Output Quality (OQ) & 4.19 (0.52) & 0.424 (0.085) & 0.526 (0.040*) & - \\
Result Demonstrability (RD) & 3.90 (0.60) & 0.226 (0.240) & 0.873 (0.001**) & - \\
Computer Self-Efficacy (CSE) & 4.15 (0.64) & 0.250 (0.216) & - & 0.364 (0.059) \\
Perception of External Control (PEC) & 3.83 (0.72) & 0.461 (0.066) & - & 0.122 (0.066) \\
Computer Anxiety (CA) & 2.83 (0.81) & -0.363 (0.123) & - & -0.337 (0.142) \\
Computer Playfulness (CP) & 3.63 (0.63) & -0.278 (0.191) & - & -0.111 (0.366) \\
Perceived Enjoyment (PE) & 3.52 (0.63) & 0.563 (0.028*) & - & 0.727 (0.004**) \\
Driving Experience (Exp) & 3.42 (5.01) & -0.495 (0.051) & -0.007 (0.492) & -0.23 (0.236) \\ \hline
\end{tabular}
\end{table*}

\begin{figure*}
    \centering
    \includegraphics[width=0.8\textwidth]{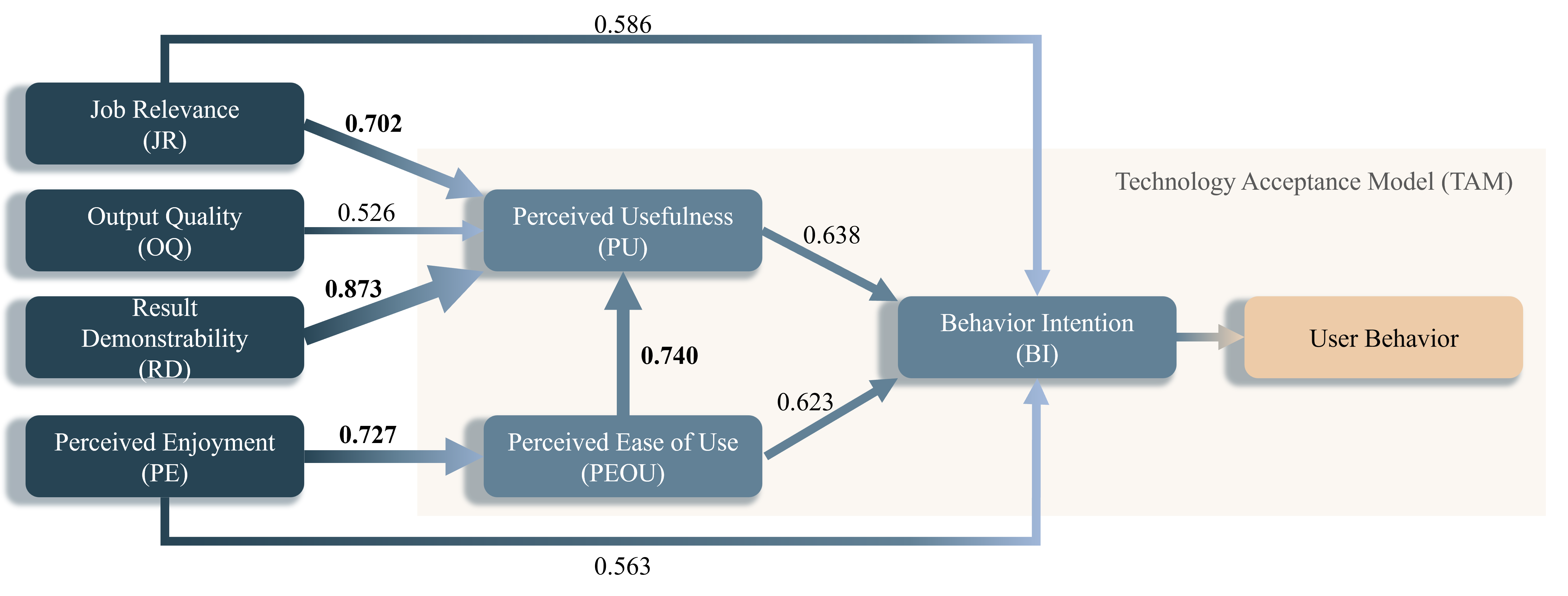}
    \caption{The result of TAM3.}
    \label{fig:tam3}
\end{figure*}

According to TAM3 analysis \cite{venkatesh2008tam3}, Pearson correlation coefficients were computed to determine the relationship between Behavioral Intention (BI) and its determinants (PU, PEOU, SN, I, JR, OQ, RD, CSE, OEC, CA, CP, PE, and Exp), between Perceived Usefulness (PU) and its determinants (PU, PEOU, SN, I, JR, OQ, RD, and Exp), and between Perceived Ease of Use (PEOU) and its determinants (PEC, CA, CP, PE, and Exp) \cite{venkatesh2008tam3, nikolopoulos2018completeTAM3}. Results are shown in table \ref{table:tam}. For correlations with BI, the positive relationships between JR and BI ($r_{p} = 0.586$, $p = 0.023$), between PE and BI ($r_{p} = 0.563$, $p = 0.028$), between PEOU and BI ($r_{p} = 0.623$, $p = 0.015$), and between PU and BI ($r_{p} = 0.638$, $p = 0.017$) were revealed, which were consistent with the TAM3 framework.
For correlations with PU, strong positive relationships between JR and PU ($r_{p}= 0.702$, $p = 0.005$), between RD and PU ($r_{p} = 0.873$, $p < 0.001$), and between PEOU and PU ($r_{p} = 0.740$, $p = 0.003$), were revealed. The positive relationship between OQ and PU was found ($r_{p} = 0.526$, $p = 0.040$). For correlations with PEOU, a strong positive relationship between PE and PEOU was found ($r_{p} = 0.727$, $p = 0.004$). The integrated result of TAM3 is demonstrated in figure \ref{fig:tam3}, in which all the significant strong correlations are shown. 

\subsubsection{Affects on Driving Behavior}
To investigate users' behaviors when interacting with the designed interface in a driving context, hybrid measures have been conducted on video analysis and vehicle telemetry data. The experiment video has been coded into three events:
\begin{itemize}
    \item Activate the Speech: The participant starts to give a verbal deictic command to the system.
    \item Pointing: The participant is pointing at a target outside.
    \item Response to the Speech: After getting the audio information about the target, the participant responds to the system.
\end{itemize}

The event segmentation was completed by using Boris. For the telemetry data, the variation in velocity has been focused on. The rate of speed change was categorized as "decreased" ($\Delta v_{min} < -0.1 mph$), "increased" ($\Delta v_{max} > 0.1 mph$), and "stable ($-0.1 mph < \Delta v < 0.1 mph$).

A hybrid clustering has been made on users' interaction behavior and change in speed by using R. The result is shown in figure \ref{fig:result}c. It is noticeable that when drivers were pointing, the velocity decreased most of the time (67\%). The result of the chi-square independence test with cross-tabulation revealed there was an association between velocity variation and drivers' behavior ($\chi^2 = 10.70$, $p = 0.03$). The percentage of deceleration with Pointing was significantly higher than the percentage of deceleration with Response to the Speech ($p < 0.05$). The percentage of declaration with Pointing was significantly higher than the percentage of stabilization with Pointing ($p < 0.05$).

Figure \ref{fig:result}d illustrates P14's driving behavior and the change in speed in time series as a case demonstration. A pattern that the driver decelerates greatly when giving verbal deictic commands and pointing, and the speed returns to stable when responding to the speech assistant can be found.

\subsubsection{Workload and Preference}
A compression analysis has been made on users' workload and preference of speech and pointer configuration, to confirm the result from the previous workshop.

\begin{figure*}
    \centering
    \includegraphics[width=1\textwidth]{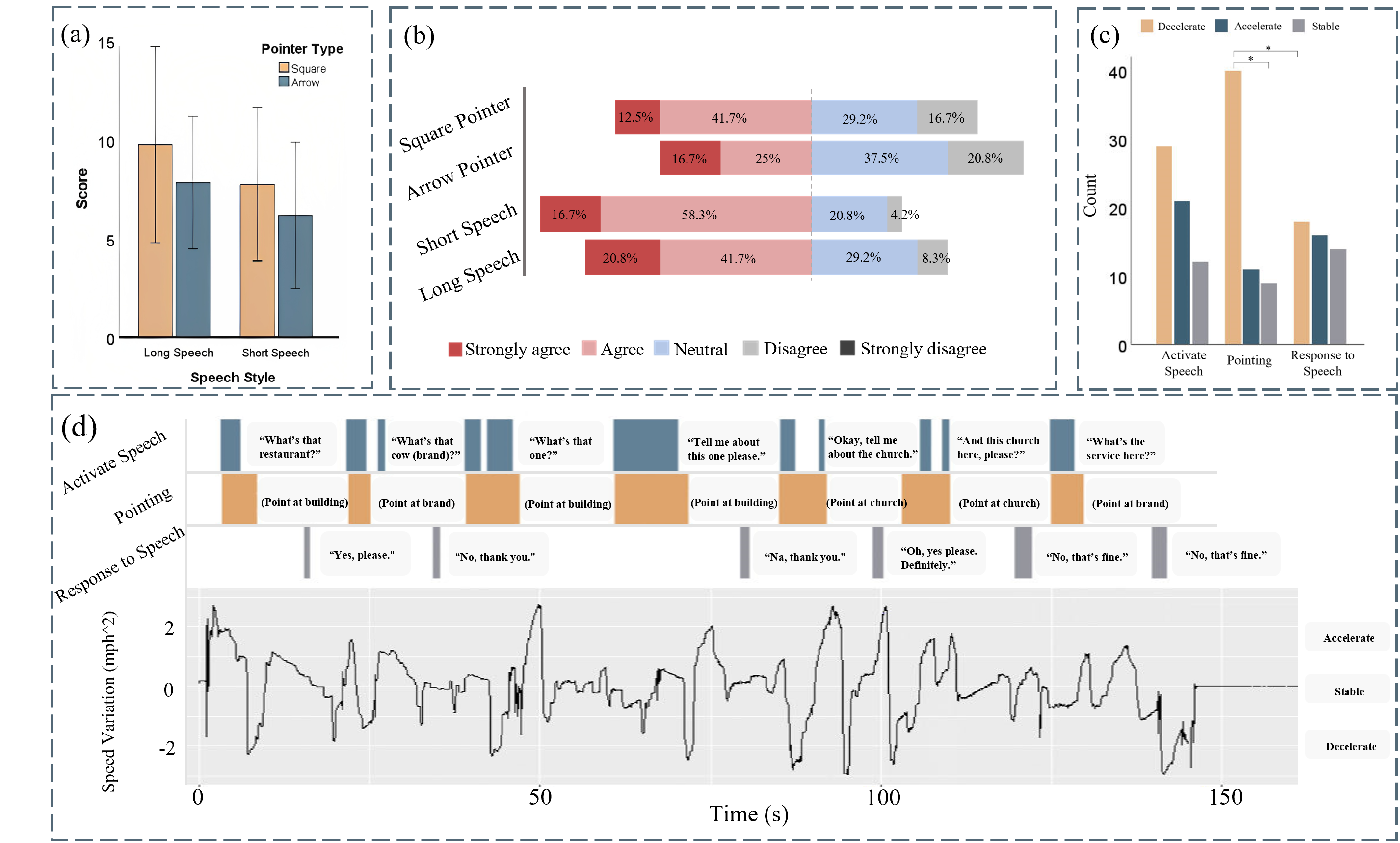}
    \caption{The result of user experiment. (a) The result of SWAT analysis. (b) Participants' preference for different visual and audio configurations. (c) The result of association analysis of users' behavior. (d) A case demonstration of P14's interaction behavior and driving speed.}
    \label{fig:result}
\end{figure*}

For the workload measurement, the results of SWAT were analyzed by ANOVA. The overall workload mean score was 7.88 ($SD = 6.26$). The analysis (see figure \ref{fig:result}a) revealed no significant effects between different types of pointers and speech styles ($F_{pointer(1,24)} = 3.00$, $p_{pointer} > 0.11$, $F_{speech(1,24)} = 4.24$, $p_{speech} > 0.06$). While using long speech, the square pointer ($M = 9.75$, $SD = 7.82$) showed a higher workload than the arrow pointer ($M = 7.83$, $SD = 5.27$). For short speech content, the square pointer ($M = 7.75$, $SD = 6.10$) showed a higher workload than the arrow pointer ($M = 6.17$, $SD = 5.83$). The workload of both pointer types in long speech was higher than in short speech. However, no significance was found for workload measurement.

For the preference measurement, participants' ratings have been analyzed (figure \ref{fig:result}b). No significance has been found between different types of pointers and speech styles ($M_{square} = 3.50$, $SD_{square} = 0.93$, $M_{arrow} = 3.38$, $SD_{arrow} = 1.01$, $Z_{pointer} = -0.62$, $p_{pointer} = 0.53$; $M_{long} = 3.57$, $SD_{long} = 0.90$, $M_{short} = 3.88$, $SD_{short} = 0.74$, $Z_{speech} = -0.42$, $p_{speech} = 0.67$), which was consistent with results from the previous workshop. 

\subsection{Interview}
During the interview, participants expressed their feelings about interacting with the design interface and provided suggestions for improvement.

As a result of transcript coding, most participants (92\%) showed positive attitudes toward the system design. The description keywords included "usefulness", "time-saving", "novel experience", etc. Participants commented that the interface was "easy to use" as they were not told how the system worked but completed their tasks without confusion. 
\begin{quote}
    P14: "Interesting. I can definitely see that it has specific uses where it would really add value".
    
    P09: "There is a new thing I've never experienced".
\end{quote}

However, some participants also showed concerns about implementing the interface on a real vehicle. P13 believed she would use the system depending on the traffic conditions. 
\begin{quote}
    P13: "I like it (the interface). And if it were in a (real) car, I would use it, but depends on how much traffic would be there. If it's in a stressful (busy) city kind of traffic, maybe I would use it less".
\end{quote}

A main reason could be the cost of mental workload and distraction. Summarized from the transcript coding, 77\% participants reported distractions when interacting with the interface. 

Some participants also mentioned the reliability of technology, such as the detection accuracy, the response delay, and the conflicts between the pointer and other HUD applications on the windshield. It was concluded that the negative impression of the interface was related to the system performance and the road situation.

\section{Discussion}
In this study, we discussed the design potential for an in-vehicle multi-modal intuitive interface, implemented the designed interface on both a driving simulator and a real autonomous vehicle, and conducted a user experiment on the simulator. Findings showed users' acceptance of the intuitive interface was significantly affected by 6 factors. The interaction behaviors had impacts on drivers' primary tasks and workload. We will also discuss the design implications for the interface on autonomous vehicles and the limitations of this study.

\subsection{Users' Acceptance to Intuitive Interface}
Association analysis of TAM3 was conducted to assess the applicability of the design in a given scenario from users' perspective and gain insight into factors that should be focused on.

The significant positive associations between \textit{perceived ease of use}, \textit{perceived usefulness} and \textit{behavioral intention} revealed the intuitive interface had both attractive features and strong functional ability, which contributed greatly to drivers' intention of using this interface. 
There was a significantly positive relation between \textit{perceived ease of use} and \textit{perceived usefulness}, which indicated the usefulness of the interface was determined by its simplicity. 
Moreover, the \textit{perceived ease of use} was significantly influenced by the \textit{perceived enjoyment}, showing the design of the intuitive interface should emphasize the entertainment of use.
Furthermore, the \textit{job relevance}, \textit{output quality}, and \textit{result demonstrability} were three major determinants that had significantly positive effects on \textit{perceived usefulness}. The effect of JR indicated the intuitive interface was highly applicable to drivers' tasks \cite{venkatesh2000theoretical}. It can be figured from the effect of OQ that participants believed the output quality was significant to the usability of the interface. The result of RD revealed the intuitive interface was tangible, observable, and communicable \cite{moore1991development}, which also had great contributions to the usefulness.

In summary, the key components for designing an in-vehicle muli-modal intuitive interface are providing joyfulness to drivers, enabling the quality of input detection and output demonstration, and the simplicity of usage. 

\subsection{Influence on Drivers' Behavior and Workload}
Even though one of the design purposes of the intuitive interface was to facilitate drivers' experience without changing their driving behavior. However, it was noticeable from the result of the hybrid analysis that there was a pattern in participants' behavior. When interacting with the interface, there was significantly more deceleration during the pointing stage. There can be two explanations. Firstly, drivers intended to slow down and observe the building as it caused his/her interest. In this case, the behavior change followed the users' interactive intention. For example, P12 was afraid the system might miss the pointed building that she deliberately drove slowly. Moreover, \cite{freehandpoint} suggested that adjusting speed is a subconscious action, and this action was not influenced by the pointing gesture itself.

On the other hand, it may caused by the distraction from the primary task, and the manifestation is unconscious deceleration. It has been suggested that measuring the speed variation was a proven approach to evaluate drivers' distraction \cite{yusoff17distractionMeasures, WANG2018236speedmeasure}. Therefore, we will also discuss the distraction and workload when using this interface.

The SWAT analysis showed the average workload was 7.88, which was a relatively small workload for performing a secondary task while driving, compared with driving only, driving with a passenger speaking, and driving with talking on a cell phone \cite{waugh2000cognitive}. The intuitive interaction has advantages in assisting in complicated primary tasks (like driving).

During the experiment interview, participants explained their subjective feelings of distraction. P15 recalled there was an \textbf{auditory distraction} when she pointed at the landmark and listened to the speech feedback simultaneously, which made her feel unsafe. 
\textbf{Visual distractions} occur when the driver looks away from the road. P17 thought he paid too much attention to the pointer showing on the windshield.
As a result of the pointing gesture, P13 mentioned she was continuously pointing to the building subconsciously, which created \textbf{physical distractions} and affected her control of the steering wheel. 
Two participants (P14, P15) mentioned there were \textbf{cognitive distractions} that they removed their attention from driving. 

Therefore, even though the deceleration action could be a natural behavior and the designed intuitive interface caused a low workload, it was still indispensable that there were various types of distractions arise, which can be potential safety hazards.

\subsection{Design Implications for Autonomous Vehicle Interface}
To explore the design and user experience of the in-vehicle multi-modal intuitive interface in the context of a driving scenario, we implemented a design interface on a simulator to conduct a user experience. It has been suggested that manual driving has more demands on in-vehicle pointing gestures compared with autonomous driving scenarios \cite{gomaa2020studyingOutsidePointMovingCar}. However, both workshop and experiment participants mentioned the autonomous vehicle could be a more suitable solution. Therefore, in the part of System Design, we also set up the interface in a real autonomous vehicle without testing it, limited by technique and ethical issues. By integrating the joint inside-outside perception, various novel functions can be developed. For instance, participants in the user experiment expected the interface to have a "point-to-go" function, that could drive to the place they pointed at by using autonomous driving. The safety issue of distraction when driving can also be tackled using autonomous vehicles.

Moreover, users' feedback in this research may provide design insights for an AV intuitive interface. For example, participants tended to decelerate when pointing to a target to have a detailed look. This feature can be implemented in an AV system, which also decelerates when detecting users' deictic gestures outside the vehicle, to provide a better user experience.

In summary, our explorations on the in-vehicle multi-modal intuitive interface provide analysis and evidence of the system usability and user experience, which can inspire future HMI development in autonomous vehicles. Our findings encourage researchers to discover the usage of intuitive interactions in vehicles, especially under a dynamic driving context.

\subsection{Limitations and Future Work}
There were several limitations in this research. Firstly, our system was realized in a Wizard-of-Oz style. The method may eliminate the impact of incomplete technology on the user experience, but may also cover the corresponding problems. In addition, our system was lack of experiment on-road using a real vehicle. There may be essential differences in experience between the simulator and the real road, regardless of the fidelity of the simulator \cite{allen2011short}. Finally, our participants' demographic is mainly distributed among young people with short-term driving experience. Elder people or drivers with extensive manual driving experience may provide different suggestions.

To improve the design of the in-vehicle intuitive interface and evaluate users' experience more effectively, this study might be improved to implement actual algorithm and exam on real road with safety constraints. More diverse demographic participants might be recruited. 

\section{Conclusion}
In this research, we explored users' needs and potential solutions and application scenarios for in-vehicle multi-modal intuitive interface through a design workshop. It has been concluded that deictic gestures, visual feedback, and speech commands are essential modalities. It was agreed that the intuitive interface has advantages in being user-friendly, having high acceptance, and low learning cost. Concerns, such as safety and technical realization, have also been raised.

With the discovery from the workshop, we implemented the designed interface on both a driving simulator and a real autonomous vehicle.
A user experiment has been conducted on the simulator to evaluate user acceptance, behavior, and experience. Findings revealed that users' acceptance of the intuitive interface is mainly influenced by the perceived usefulness, perceived ease of use, job relevance, output quality, result demonstrability, and perceived enjoyment. Drivers have been proven to have less workload but distractions were reported during the interview. Participants suggested the in-vehicle multi-modal intuitive interface is a novel and attractive design that they would like to use in real life. 

The intuitive interface in this research was a preliminary exploration. There are still spaces for various technical and functional improvements. 
Our findings may provide insights into both manual driving and autonomous vehicle interface design.

\begin{acks}
We want to thank all the participants in this study.
\end{acks}

\bibliographystyle{ACM-Reference-Format}
\bibliography{sample-base}

\onecolumn
\appendix

\section{The semi-structured interview of the design workshop}
\label{ap:workshop}
\begin{enumerate}
    \item What are the advantages and disadvantages of using intuitive interaction for in-vehicle interfaces? (compare to non-intuitive interactions)
    \item (Fellowed the feedback of question 1) What factors make your interface highly acceptable to users?
    \item How would interactions be different while driving versus while parking?
    \item What kinds of visual instructions are better?
    \item What kinds of voice feedback are better? (Simple or rich, long or short?)
    \item Do you have any other ideas?   
\end{enumerate}

\section{The semi-structured interview of the user experiment}
\label{ap:exp}
\begin{enumerate}
    \item How do you like this interface? What are the advantages and disadvantages of this interface? Why?
    \item There are different variables between trails (pointer types and speech content styles), did they feel the same? Which one do you prefer and why?
    \item Did you feel distracted when using it? How?
    \item Do you have any other feelings and thoughts?
\end{enumerate}

\end{document}